\documentclass[11pt,a4paper]{article}

\usepackage{amssymb}
\usepackage{graphicx}
\usepackage{bm}

\usepackage{mathrsfs}
\usepackage{cite}

\begin{document}

\title{Algebraic structure of the renormalization group in the renormalizable QFT theories}

\author{
A.L.Kataev${}^a$\footnote{kataev@ms2.inr.ac.ru} and K.V.Stepanyantz${}^b$\footnote{stepan@m9com.ru}\\
\\
${}^a${\small{\em Institute for Nuclear Research of the Russian Academy of Science,}}\\
{\small {\em prospekt 60-letiya Oktyabrya 7a, Moscow, 117312, Russia}},\\
${}^b${\small{\em Moscow State University, Faculty of Physics,}}\\ {\small{\em Department of Theoretical Physics,}}\\
{\small{\em Leninskie Gory, Moscow, 119991, Russia}}\\
}

\maketitle

\begin{abstract}
We consider the group formed by finite renormalizations as an infinite-dimensional Lie group. It is demonstrated that for the finite renormalization of the gauge coupling constant its generators $\hat L_n$ with $n\ge 1$ satisfy the commutation relations of the Witt algebra and, therefore, form its subalgebra. The commutation relations are also written for the more general case when finite renormalizations are made for both the coupling constant and matter fields. We also construct the generator of the Abelian subgroup corresponding to the changes of the renormalization scale. The explicit expressions for the renormalization group generators are written in the case when they act on the $\beta$-function and the anomalous dimension. It is explained how the finite changes of these functions under the finite renormalizations can be obtained with the help of the exponential map.
\end{abstract}

\vspace*{-14.0cm}

\begin{flushright}
INR-TH-2024-003
\end{flushright}

\vspace*{13.0cm}

\section{Introduction}
\hspace*{\parindent}

It is well known that quantum corrections in most field theory models are divergent. In renormalizable theories these divergences can be absorbed into the redefinition of various constants and fields. However, the renormalizations constants are not uniquely defined \cite{Bogolyubov:1980nc,Collins:1984xc,Stevenson:2022gcv}. In each order of the perturbation theory there are some arbitrary constants which determine a specific subtraction scheme. According to \cite{Kallen:1954jfh,Vladimirov:1975mx,Vladimirov:1979my,Vladimirov:1979ib} various renormalization prescriptions are related by finite renormalizations first introduced in \cite{Kallen:1954jfh}. These finite renormalization form a Lie group included into the renormalization group \cite{StueckelbergdeBreidenbach:1952pwl,Gell-Mann:1954yli,Bogolyubov1,Bogolyubov:1956gh}. From the group structure of finite renormalization it is possible to obtain some important results. In particular, in the case of using the cut-off type regularizations (e.g., the higher covariant derivative regularization \cite{Slavnov:1971aw,Slavnov:1972sq,Slavnov:1977zf}) the renormalization group functions (RGFs) do not depend on the $\ln\Lambda/\mu$, where $\Lambda$ is an ultraviolet cut-off and $\mu$ is a renormalization point. For dimensional regularization \cite{'tHooft:1972fi,Bollini:1972ui,Ashmore:1972uj,Cicuta:1972jf} or reduction \cite{Siegel:1979wq} in $D=4-\varepsilon$ dimensions the four-dimensional RGFs do not depend on $\varepsilon^{-1}$. This, in turn, allows relating the coefficients at higher $\varepsilon$-poles or/and logarithms to the coefficients of RGFs \cite{tHooft:1973mfk}, see also \cite{Kazakov:2008tr}. The explicit form of the corresponding expressions can be found in \cite{Derkachev:2017nhd,Meshcheriakov:2022tyi} for the cut-off type regularizations and in \cite{Ivanov:2017ekx,Ivanov:2018pga,Meshcheriakov:2023fmk} for various versions of the dimensional regularization.

The group structure of the renormalization group was studied in literature, see, e.g., \cite{Osborn:1991gm,Kovalev:1996ze,Shirkov:1999hj,Kovalev:2008ht,Kamenshchik:2020yyn}, but (to the best of our knowledge) some details have not still been investigated. In this paper we construct the (infinite-dimensional) Lie algebra corresponding to the renormalization group and its one-dimensional subalgebra, corresponding to the changes of the renormalization scale. In particular, we demonstrate that the generators of the coupling constant  finite renormalizations form a certain subalgebra (generated by the operators $\hat L_n$ with $n\ge 1$) of the Witt algebra \cite{Cartan} (whose central extension is widely known as the Virasoro algebra \cite{Virasoro:1969zu}). The corresponding Lie group is then obtained with the help of the exponential map.

\section{The renormalization group preliminaries}
\hspace*{\parindent}

The renormalization of the coupling constant\footnote{For simplicity, we will consider a theory with a single coupling constant.} can be made by splitting the bare coupling constant $\alpha_0$ into the renormalized coupling constant $\alpha$ and the counterterms,

\begin{equation}
\alpha_0 = \alpha_0\Big(\alpha(\mu),\ln\Lambda/\mu\Big),
\end{equation}

\noindent
where $\Lambda$ is a dimensionful regularization parameter, and $\mu$ is a renormalization point. For renormalizing fields, masses, etc. we again split the bare values into the product of the renormalized value and the relevant renormalization constant, e.g.,

\begin{equation}
\varphi = Z(\alpha,\ln\Lambda/\mu) \varphi_{R},
\end{equation}

\noindent
where, for simplicity, we omitted all possible indices. (Note that in the case of using dimensional technique divergences appear as poles in $\varepsilon\equiv 4-D$, but even in this case it is possible to introduce the regularization parameter $\Lambda$ different from the renormalization point $\mu$, see, e.g., \cite{Meshcheriakov:2023fmk,Aleshin:2019yqj}.)

The divergences in the bare coupling constant and fields can conveniently be encoded in RGFs, namely, in the $\beta$-function and anomalous dimensions. In our notations they are defined by the equations

\begin{equation}\label{RGFs_Definitions}
\beta(\alpha) \equiv \frac{d\alpha(\mu)}{d\ln\mu}\bigg|_{\alpha_0=\mbox{\scriptsize const}}; \qquad \gamma(\alpha) \equiv \frac{d}{d\ln\mu} \ln Z(\alpha,\ln\Lambda/\mu)\bigg|_{\alpha_0=\mbox{\scriptsize const}}.
\end{equation}

The renormalized ($n$-point) Green's function $G_R$ is related to the corresponding bare function $G$ be the equation

\begin{equation}\label{Gamma_Renormalized}
G_R\Big(\alpha(\mu),p_1/\mu,\ldots, p_{n}/\mu\Big) = Z_G(\alpha,\ln\Lambda/\mu) G\Big(\alpha_0,p_1/\Lambda,\ldots,p_{n}/\Lambda\Big),
\end{equation}

\noindent
where the external momenta $p_i$ satisfy the constraint $p_1+\ldots+p_n=0$, and the Lorentz indices were omitted or simplicity. Note that, for simplicity, we consider the limit in which all momenta are large, so that it is possible to omit various masses. The renormalization constant $Z_G$ can certainly be expressed in terms of the renormalization constants for various fields of the theory. The renormalized Green's function satisfies the Ovsyannikov--Callan--Symanzik (OCS) equation \cite{Ovsyannikov:1956fa,Callan:1970yg,Symanzik:1970rt,Symanzik:1971vw}

\begin{equation}\label{OCS}
0 = \Big(\frac{\partial}{\partial\ln\mu} + \beta(\alpha) \frac{\partial}{\partial\alpha} - \gamma_G(\alpha)\Big) G_R,
\end{equation}

\noindent
where the anomalous dimension of $G$ is defined similarly to Eq. (\ref{RGFs_Definitions})

\begin{equation}
\gamma_G(\alpha) \equiv \frac{d}{d\ln\mu} \ln Z_G(\alpha,\ln\Lambda/\mu)\bigg|_{\alpha_0=\mbox{\scriptsize const}}.
\end{equation}

However, the renormalization constants are not defined uniquely \cite{Kallen:1954jfh}. For example, the renormalized coupling constant is defined up to the finite renormalizations of the form

\begin{equation}
\alpha'(\mu') = \alpha'(\alpha,\ln\mu'/\mu).
\end{equation}

\noindent
Similarly, it is possible to redefine the renormalization constant for the matter fields making the finite renormalization

\begin{equation}\label{Finite_Z}
Z'(\alpha',\ln\Lambda/\mu') = z(\alpha,\ln\mu'/\mu) Z(\alpha,\ln\Lambda/\mu).
\end{equation}

\noindent
In general, even for $\mu'=\mu$ these finite renormalizations are not trivial, $\alpha'(\alpha) \equiv \alpha'(\alpha,0)\ne \alpha$ and $z(\alpha) \equiv z(\alpha,0) \ne 1$. In this case under the finite renormalizations RGFs change as \cite{Vladimirov:1979ib,Vladimirov:1979my}

\begin{equation}\label{RGFs_Change}
\beta'(\alpha') = \frac{d\alpha'}{d\alpha} \beta(\alpha);\qquad \gamma'(\alpha') = \frac{d\ln z}{d\alpha} \beta(\alpha) + \gamma(\alpha).
\end{equation}

\noindent
The perturbative expansions of RGFs can be written as

\begin{equation}
\beta(\alpha) = \sum\limits_{n=1}^\infty \beta_n \alpha^{n+1};\qquad \gamma(\alpha) = \sum\limits_{n=1}^\infty \gamma_n \alpha^n,
\end{equation}

\noindent
where the index $n$ numerates an order of the perturbation theory. From these equations it is possible to see that the coefficients of the $\beta$-function and of the anomalous dimension in general change starting from the three- and two-loop order, respectively. In other words, in general, $\beta_n'\ne \beta_n$ for $n\ge 3$, and $\gamma_n'\ne\gamma_n$ for $n\ge 2$. The transformations under which RGFs remain invariant in all orders are reduced to the change of the renormalization point.

The finite renormalizations determined by the functions $\alpha'(\alpha,\ln\mu'/\mu)$ and $z(\alpha,\ln\mu'/\mu)$ form the group, see, e.g., \cite{Bogolyubov:1980nc,Collins:1984xc}. Its structure will be investigated in what follows.

\section{Finite renormalization of the gauge coupling constant}

\subsection{Generators and their commutation relations}
\hspace*{\parindent}

Let us now specify the group structure of the renormalization group. First, for simplicity, we consider the finite renormalizations for which $\mu'=\mu$ and $z(\alpha)=1$. They can be presented in the form $\alpha\to \alpha'(\alpha)$.

For any Lie group ${\cal G}$ in a certain vicinity of the identity element 1 the group element $\hat\omega \in {\cal G}$ can be presented as the exponential of the corresponding Lie algebra ${\cal A}$ element $\hat a \in {\cal A}$,

\begin{equation}
\hat\omega = \exp(\hat a),
\end{equation}

\noindent
see, e.g., \cite{Isaev:2018xcg}. To describe the group structure of the renormalization group (in a neighborhood of the identity element), it is sufficient to construct its Lie algebra. This algebra is determined by the commutation relations of its generators. These generators are obtained by considering the infinitesimal transformations. In the considered case they correspond to the infinitesimal finite renormalization of the coupling constant, which can be presented as the series

\begin{equation}\label{Alpha_Expansion}
\delta\alpha = -\sum\limits_{n=1}^\infty a_n \alpha^{n+1} \equiv \sum\limits_{n=1}^\infty a_n \hat L_n \alpha \equiv \hat a \alpha,
\end{equation}

\noindent
where $a_n$ are arbitrary (real) constants parameterizing the finite renormalizations of charge. By definition, $\delta\alpha$ includes only terms linear in the parameters $a_n$. The operators $\hat L_n$ are the generators of the renormalization group, where $n\ge 1$ due to using of the perturbation theory. The corresponding finite transformations are obtained with the help of the exponential map, $\alpha' = \hat\omega\alpha = \exp(\hat a)\alpha$. The explicit examples will be considered in Sections \ref{Subsection_Finite_Alpha} and \ref{Subsection_Finite}. In this case the expression $\alpha'-\alpha$ contains not only terms linear in the parameters $a_n$, but also their higher powers,

\begin{equation}
\alpha'-\alpha =\hat\omega\alpha - \alpha = \exp\Big(\sum\limits_{n=1}^\infty a_n \hat L_n\Big) \alpha - \alpha = \delta\alpha + O(a^2),
\end{equation}

\noindent
where (here and in what follows) $O(a^2)$ denotes all terms with higher powers of $a_n$ (starting from the quadratic ones). They include all terms beyond the linear approximation. Note that, in general, the $\beta$-function changes under these transformations \cite{Stevenson:2022gcv}, and the corresponding change of its coefficients can be expressed in terms of the parameters $a_n$.

From Eq. (\ref{Alpha_Expansion}) we see that (for an arbitrary positive integer $n$) the generators of the renormalization group in the explicit form are written as

\begin{equation}
\hat L_n\alpha \equiv - \alpha^{n+1}.
\end{equation}

\noindent
If these generators act on an arbitrary function of $\alpha$, then from this equation we easily obtain that

\begin{equation}\label{Ln_Explicit}
\hat L_n = - \alpha^{n+1} \frac{d}{d\alpha}.
\end{equation}

These operators satisfy the commutation relations of the Witt algebra\footnote{ A relation between the renormalization group and the Witt algebra may also  be established from the material
of the task 2.2.15 of Ref. \cite{Isaev:2018xcg}. We are grateful to A.P.Isaev for this comment.} \cite{Cartan}
 because for an arbitrary function $f(\alpha)$ we have

\begin{eqnarray}\label{Witt}
&& [\hat L_n,\hat L_m] f(\alpha) = - \hat L_n \alpha^{m+1} \frac{df(\alpha)}{d\alpha} + \hat L_m \alpha^{n+1} \frac{df(\alpha)}{d\alpha} =  \alpha^{n+1} \frac{d}{d\alpha}\Big(\alpha^{m+1}\frac{df(\alpha)}{d\alpha}\Big)
\qquad\nonumber\\
&& - \alpha^{m+1} \frac{d}{d\alpha}\Big(\alpha^{n+1}\frac{df(\alpha)}{d\alpha}\Big) = (m-n) \alpha^{n+m+1} \frac{df(\alpha)}{d\alpha} = (n-m) \hat L_{n+m} f(\alpha).\vphantom{\Big(}\qquad
\end{eqnarray}

\noindent
However, in the Witt algebra $n$ is an arbitrary positive integer, while in the case under consideration $n\ge 1$. Therefore, we conclude that finite renormalizations form a subalgebra of the Witt algebra generated by all $\hat L_n$ with $n\ge 1$. Note that the well-known Virasoro algebra \cite{Virasoro:1969zu} widely used in string theory \cite{Green:1987sp} is the central extension of Witt algebra. However, the difference between the Witt and Virasoro algebras is essential only for $m+n=0$, while in our case $m,n\ge 1$. From Eq. (\ref{Witt}) we in particular see that the renormalization group is non-Abelian in agreement with the consideration of \cite{Goriachuk:2020wyn} based on different arguments.

\subsection{Some representations of the renormalization group}
\hspace*{\parindent}

Due to Eq. (\ref{Gamma_Renormalized}) under the finite renormalization of the coupling constant the Green's functions change according to the equation

\begin{equation}\label{Green_Change}
G_R'\Big(\alpha'(\mu'),p_1/\mu',\ldots,p_n/\mu'\Big) = z_G(\alpha,\ln\mu'/\mu) G_R\Big(\alpha(\mu),p_1/\mu,\ldots,p_n/\mu\Big).
\end{equation}

\noindent
Let us first consider the case $\mu'=\mu$ and $z_G(\alpha)\equiv z_G(\alpha,0)=1$. Then it is possible to present the infinitesimal change of the Green's function in the form

\begin{equation}\label{Delta_G}
G_R'(\alpha,p/\mu) - G_R(\alpha,p/\mu) =  \sum\limits_{n=1}^\infty a_n T_1(\hat L_n) G_R(\alpha,p/\mu) + O(a^2) \equiv \delta G_R + O(a^2),
\end{equation}

\noindent
where $T_1(\hat L_n)$ are the generators in the considered representation, and, for simplicity, $p/\mu$ denotes the dependence on all momenta. Note that, by definition, $\delta G_R$ includes only terms linear in the parameters $a_n$. Taking into account that (for $z_G=1$)

\begin{eqnarray}
&& G_R(\alpha,p/\mu) = G_R'(\alpha',p/\mu) = G_R'(\alpha,p/\mu) + \delta\alpha \frac{\partial}{\partial\alpha} G_R(\alpha,p/\mu) + O(a^2)\qquad
\nonumber\\
&& = G_R'(\alpha,p/\mu) - \sum\limits_{n=1}^\infty a_n \alpha^{n+1} \frac{\partial}{\partial\alpha} G_R(\alpha,p/\mu) + O(a^2)\qquad
\end{eqnarray}

\noindent
we see that acting on the Green's functions the generators take the form

\begin{equation}\label{Green_Representation}
T_1(\hat L_n) G_R = \alpha^{n+1} \frac{\partial}{\partial\alpha} G_R.
\end{equation}

\noindent
Note that in Eq. (\ref{Delta_G}) the argument of the function $G_R'$ is the same as of the function $G_R$, so that $T_1(\hat L_n)\alpha = 0$. Therefore, a different sign in Eq. (\ref{Green_Representation}) in comparison with Eq. (\ref{Ln_Explicit}) is really needed,

\begin{eqnarray}
&& [T_1(\hat L_n),T_1(\hat L_m)] G_R = T_1(\hat L_n)\, \alpha^{m+1}\frac{\partial}{\partial\alpha} G_R - (n\leftrightarrow m)
\nonumber\\
&& = \alpha^{m+1}\frac{\partial}{\partial\alpha} \Big(\alpha^{n+1}\frac{\partial}{\partial\alpha} G_R\Big) - (n\leftrightarrow m) = (n-m) \alpha^{n+m+1} \frac{\partial}{\partial\alpha} G_R
\nonumber\\
&& = (n-m) T_1(\hat L_{n+m})\, G_R. \vphantom{\frac{\partial}{\partial\alpha}}
\end{eqnarray}

Similarly, it is possible to consider a different representation of the renormalization group in which its generators act on the $\beta$-function. Indeed, it is known that RGFs nontrivially transform under the finite renormalizations \cite{Vladimirov:1979ib,Vladimirov:1979my}. Namely, the $\beta$-function changes according to the first equation in (\ref{RGFs_Change}). Making the transformations (\ref{Alpha_Expansion}) we can find the explicit form of the generators $T_2(\hat L_n)$ in this representation. Really, if we consider a finite renormalization (\ref{Alpha_Expansion}), then from the first equation in (\ref{RGFs_Change}) we conclude that (in the linear order in the parameters $a_n$) the $\beta$-function changes as

\begin{equation}
\beta'\Big(\alpha - \sum\limits_{n=1}^\infty a_n \alpha^{n+1}\Big) = \Big(1 - \sum\limits_{n=1}^\infty a_n (n+1) \alpha^n\Big) \beta(\alpha) + O(a^2).
\end{equation}

\noindent
Expanding the left hand side into a Taylor series and keeping only terms linear in $a_n$ we obtain the infinitesimal change of the $\beta$-function,

\begin{eqnarray}\label{Delta_Beta}
&& \beta'(\alpha) - \beta(\alpha) = \sum\limits_{n=1}^\infty a_n\Big(\alpha^{n+1} \frac{d\beta(\alpha)}{d\alpha} - (n+1)\alpha^n\beta(\alpha)\Big) + O(a^2) \qquad\nonumber\\
&& \equiv \sum\limits_{n=1}^\infty a_n T_2(\hat L_n) \beta(\alpha) + O(a^2) \equiv \delta\beta(\alpha) + O(a^2),
\end{eqnarray}

\noindent
where the operators $T_2(\hat L_n)$ are the generators of considered representation of the renormalization group,

\begin{equation}\label{T_Beta_Representation}
T_2(\hat L_n)\,\beta(\alpha) = \Big(\alpha^{n+1} \frac{d}{d\alpha} - (n+1)\alpha^n\Big)\beta(\alpha).
\end{equation}

\noindent
Note that they act only on the $\beta$-function and do not act on $\alpha$, because the arguments of $\beta'$ and $\beta$ in Eq. (\ref{Delta_Beta}) are the same. Therefore, in the considered representation $T_2(\hat L_n)\alpha = 0$. Again, the commutation relations of the Witt algebra can easily be verified,

\begin{eqnarray}
&& [T_2(\hat L_n),T_2(\hat L_m)] \beta(\alpha) = T_2(\hat L_n) \Big(\alpha^{m+1}\frac{d}{d\alpha} - (m+1)\alpha^m \Big)\beta(\alpha) - (n\leftrightarrow m)\vphantom{\frac{1}{2}}\nonumber\\
&& = \Big(\alpha^{m+1}\frac{d}{d\alpha} - (m+1)\alpha^m \Big) \Big(\alpha^{n+1}\frac{d}{d\alpha} - (n+1)\alpha^n \Big) \beta(\alpha) - (n\leftrightarrow m)\nonumber\\
&& = (n-m) \Big(\alpha^{n+m+1} \frac{d}{d\alpha} - (n+m+1)\alpha^{n+m}\Big)\beta(\alpha) = (n-m) T_2(\hat L_{n+m})\beta(\alpha),\qquad
\end{eqnarray}

\noindent
where we took into account that

\begin{equation}
T_2(\hat L_n) \Big(\alpha^{m+1}\frac{d}{d\alpha} - (m+1)\alpha^m \Big)\beta(\alpha) = \Big(\alpha^{m+1}\frac{d}{d\alpha} - (m+1)\alpha^m \Big)T_2(\hat L_n) \beta(\alpha).
\end{equation}

Thus, we explicitly constructed the generators of the group formed by finite renormalizations of the coupling constant in two different representation, in which they act either on the Green's functions, or on the $\beta$-function. Certainly, in both cases they satisfy the commutation relations of the subalgebra of the Witt algebra formed by its generators with a positive indices.

\subsection{Non-infinitesimal finite renormalizations of the coupling constant}
\hspace*{\parindent}\label{Subsection_Finite_Alpha}

So far we considered infinitesimal finite renomalizations. Here we will take into account the terms of the higher orders in the parameters $a_n$ using the general arguments based on group theory. Namely, for obtaining finite transformations it is necessary to use the exponential map. Then the expression for the new gauge coupling constant can be presented as

\begin{equation}\label{Alpha_New}
\alpha' = \exp\Big(\sum\limits_{n=1}^\infty a_n \hat L_n \Big) \alpha = \exp\Big(-\sum\limits_{n=1}^\infty a_n\alpha^{n+1} \frac{d}{d\alpha}\Big) \alpha.
\end{equation}

\noindent
Similarly, the expression for $\beta$-function in the new renormalization scheme is written in the form

\begin{equation}\label{Beta_New}
\beta'(\alpha) = \exp\Big(\sum\limits_{n=1}^\infty a_n T_2(\hat L_n)\Big) \beta(\alpha) = \exp\Big(\sum\limits_{n=1}^\infty a_n \Big(\alpha^{n+1}\frac{d}{d\alpha}-(n+1)\alpha^n\Big)\Big) \beta(\alpha),
\end{equation}

\noindent
where we took Eq. (\ref{T_Beta_Representation}) into account. The expressions (\ref{Alpha_New}) and (\ref{Beta_New}) are exact in all orders of the perturbation theory. However, it is expedient to verify them in the lowest approximations. Then the new gauge coupling constant (\ref{Alpha_New}) is written as

\begin{eqnarray}\label{Alpha_Prime}
&& \alpha' = \alpha - a_1 \alpha^2 + (-a_2+a_1^2)\alpha^3 + \Big(-a_3+\frac{5}{2}a_1 a_2 - a_1^3\Big)\alpha^4 + O(\alpha^5) \nonumber\\
&& \equiv \alpha + k_1\alpha^2 + k_2\alpha^3 + k_3\alpha^4 + O(\alpha^5).\qquad
\end{eqnarray}

\noindent
Similarly, expanding the right hand side of Eq. (\ref{Beta_New}) we obtain that the new $\beta$-function in the lowest orders (up to and including the four-loop approximation) is given by the expression

\begin{eqnarray}
&& \beta'(\alpha) = \beta_1\alpha^2 + \beta_2\alpha^3 + (\beta_3 + a_1\beta_2 - a_2\beta_1)\alpha^4 + (\beta_4 + 2a_1\beta_3 - 2 a_3\beta_1 \nonumber\\
&& + a_1^2\beta_2- a_1a_2\beta_1)\alpha^5 + O(\alpha^6) \equiv \beta_1' \alpha^2 + \beta_2'\alpha^3 + \beta_3'\alpha^4 + \beta_4'\alpha^5 + O(\alpha^6). \qquad\vphantom{\Big(}
\end{eqnarray}

\noindent
From this equation we see that (as well known) the first two coefficients of the $\beta$-function do not change, $\beta_1'=\beta_1$ and $\beta_2'=\beta_2$. Expressing the coefficients $a_n$ in terms of $k_n$ with the help of Eq. (\ref{Alpha_Prime}) the transformations of the $\beta$-function next coefficients can be presented as

\begin{eqnarray}
&& \beta_3' = \beta_3 - k_1\beta_2 + k_2\beta_1 - k_1^2\beta_1;\nonumber\\
&& \beta_4' = \beta_4 - 2k_1\beta_3 + 2 k_3\beta_1 - 6k_1k_2\beta_1 + 4k_1^3\beta_1 + k_1^2\beta_2. \qquad\vphantom{\Big(}
\end{eqnarray}

\noindent
These equations can equivalently be rewritten in the form

\begin{eqnarray}
&& k_2 = \frac{\beta_3'-\beta_3}{\beta_1} + \frac{k_1\beta_2}{\beta_1} + k_1^2;\nonumber\\
&& k_3 = \frac{\beta_4'-\beta_4}{2\beta_1} + \frac{k_1(3\beta_3'-2\beta_3)}{\beta_1} + \frac{5k_1^2 \beta_2}{2\beta_1} + k_1^3,
\end{eqnarray}

\noindent
in which they exactly coincide with the expressions presented in \cite{Kataev:1988sq}, which were derived directly from the first equation in (\ref{RGFs_Change}). Therefore, the large finite renormalization can also be constructed using the group theory technique, namely, with the help of the exponential map and using the explicit expressions for the generators in the relevant representations.

\subsection{A subgroup corresponding to the change of the renormalization scale}
\hspace*{\parindent}

Now let us consider a particular case of finite renormalizations which correspond to the change of the renormalization point $\mu$. Namely, we make the transformation

\begin{equation}\label{Mu_Change}
\alpha(\mu) \to \alpha(\mu')\equiv \alpha'(\mu).
\end{equation}

\noindent
As a parameter of these transformations, it is convenient to choose

\begin{equation}
t = \ln\frac{\mu'}{\mu}.
\end{equation}

\noindent
As earlier, first, we consider the infinitesimal transformations for which $\mu'$ is close to $\mu$ (or, equivalently, the parameter $t$ is small). Then integrating the first equation in (\ref{RGFs_Definitions}) from $\mu$ to $\mu'$ (in the linear order in $t$) we obtain

\begin{eqnarray}\label{Delta_Alpha_Rescaling}
&& \alpha'-\alpha = \delta\alpha + O(t^2) = \beta(\alpha) t + O(t^2) = t\sum\limits_{n=1}^\infty \beta_n \alpha^{n+1} + O(t^2) \qquad\nonumber\\
&& = - t \sum\limits_{n=1}^\infty \beta_n \hat L_n\alpha + O(t^2),
\end{eqnarray}

\noindent
where, by definition, $\delta\alpha$ is linear in $t$. Note that for the infinitesimal transformations the right hand side contains only terms of the first order in $t$. The terms with higher powers of $t$ will be discussed below. Accoding to Eq. (\ref{Delta_Alpha_Rescaling}), the transformation (\ref{Mu_Change}) can be considered as a particular case of the finite renormalization (\ref{Alpha_Expansion}) with $a_n = -t\beta_n$. Evidently, the corresponding generator is written as

\begin{equation}\label{L_Operator}
\hat L = - \sum\limits_{n=1}^\infty \beta_n \hat L_n.
\end{equation}

\noindent
The subgroup (\ref{Mu_Change}) of the renormalization group is Abelian \cite{Brodsky:1992pq,Brodsky:2012ms}, because it contains the only generator $\hat L$, which certainly commutes with itself. The explicit form of the generator $\hat L$ acting on $\alpha$ can be found from Eqs. (\ref{Ln_Explicit}) and (\ref{L_Operator}),

\begin{equation}
\hat L = \sum\limits_{n=1}^\infty \beta_n \alpha^{n+1} \frac{d}{d\alpha} = \beta(\alpha) \frac{d}{d\alpha}.
\end{equation}

\noindent
Therefore, under the infinitesimal renormalization group transformation generated by the operator $\hat L$ the coupling constant changes as

\begin{equation}
\alpha'(\mu) = \alpha(\mu) + t\hat L\alpha + O(t^2) = \Big(1 + t \beta(\alpha) \frac{d}{d\alpha} + O(t^2)\Big)\alpha.
\end{equation}

\noindent
This equation allows constructing the corresponding finite transformations with the help of the exponential map,

\begin{equation}
\alpha(\mu') \equiv \alpha'(\mu) = \exp\Big( t \beta(\alpha)\frac{d}{d\alpha} \Big)\alpha = \exp\Big( \ln\frac{\mu'}{\mu} \beta(\alpha)\frac{d}{d\alpha} \Big)\alpha.
\end{equation}

\noindent
Evidently, a similar equation is valid for an arbitrary function $f(\alpha)$,

\begin{equation}\label{Evolution}
f(\alpha') = \exp\Big(t \beta(\alpha) \frac{d}{d\alpha}\Big) f(\alpha),
\end{equation}

\noindent
where the value of $t$ is no longer considered small. This equation is in agreement with similar equations used earlier in \cite{Groote:2001im,Mikhailov:2004iq,Kataev:2014jba,Derkachev:2017nhd} and is obtained from the equation

\begin{equation}
\exp\Big(t \beta(\alpha) \frac{d}{d\alpha}\Big) \Big(f(\alpha) g(\alpha)\Big) = \exp\Big(t \beta(\alpha) \frac{d}{d\alpha}\Big) f(\alpha) \cdot \exp\Big(t \beta(\alpha) \frac{d}{d\alpha}\Big) g(\alpha),
\end{equation}

\noindent
which can be derived with the help of the general Leibniz product rule.

Under the considered transformations the Green's functions (with $z_G=1$) change as

\begin{eqnarray}
&& T_1(\hat L)\, G_R = - \sum\limits_{n=1}^\infty \beta_n T_1(\hat L_n)\, G_R = - \sum\limits_{n=1}^\infty \beta_n \alpha^{n+1}\frac{\partial}{\partial\alpha} G_R = -\beta(\alpha) \frac{\partial}{\partial\alpha} G_R
\qquad\nonumber\\
&& = \frac{\partial}{\partial \ln\mu} G_R,
\end{eqnarray}

\noindent
where we took into account Eq. (\ref{OCS}). Therefore, the considered transformation is really a change of the renormalization scale. The $\beta$-function in this case does not change because

\begin{eqnarray}
&& T_2(\hat L)\, \beta(\alpha) = - \sum\limits_{n=1}^\infty \beta_n T_2(\hat L_n)\, \beta(\alpha) = - \sum\limits_{n=1}^\infty \beta_n \Big(\alpha^{n+1}\frac{d}{d\alpha} - (n+1)\alpha^n\Big) \beta(\alpha) \qquad\nonumber\\
&& = -\Big(\beta(\alpha)\frac{d}{d\alpha} - \frac{d\beta(\alpha)}{d\alpha}\Big) \beta(\alpha) = 0.\qquad
\end{eqnarray}

Therefore, the changes of the renormalization scale can be viewed as an Abelian subgroup of the non-Abelian group formed by general finite renormalizations of the coupling constant under which the $\beta$-function remains invariant.

\section{Finite renormalizations that include matter renormalization}

\subsection{Commutation relations for the group of finite renormalizations}
\hspace*{\parindent}

Let us now investigate the finite renormalization that include the renormalization of the matter fields and masses. Again, let us consider finite renormalizations with $\mu'=\mu$. They are determined by the functions $\alpha'(\alpha)$ and $z(\alpha)$ defined earlier. In particular, $z(\alpha) = z(\alpha,0)$, where the function $z(\alpha,\ln\mu'/\mu)$ was introduced in Eq. (\ref{Finite_Z}). For the infinitesimal finite renormalizations we present $z(\alpha)$ in the form

\begin{equation}\label{Z_Expansion}
z(\alpha) = 1 - \sum\limits_{n=1}^\infty z_n \alpha^n + O(az,z^2),
\end{equation}

\noindent
where the summation index $n$ again indicates the order of the perturbation theory, and $O(az,z^2)$ denotes the higher order terms in $z_n$ and $a_n$. According to Eq. (\ref{Finite_Z}), under this finite renormalization the renormalized fields change as

\begin{eqnarray}
&&\varphi_R' = z^{-1}(\alpha)\varphi_R = \Big(1 + \sum\limits_{n=1}^\infty z_n \alpha^n + O(az,z^2)\Big)\varphi_R\nonumber\\
&&\equiv \varphi_R + \sum\limits_{n=1}^\infty z_n \hat G_n \varphi_R + O(az,z^2),
\end{eqnarray}

\noindent
where the operators

\begin{equation}
\hat G_n \varphi_R \equiv \alpha^n \varphi_R
\end{equation}

\noindent
generate the infinitesimal finite renormalizations of the matter fields. It is easy to see that they commute with each other,

\begin{equation}
[\hat G_n,\hat G_m]\varphi_R = \Big(\alpha^{n+m} - (m\leftrightarrow n)\Big)\varphi_R = 0,
\end{equation}

\noindent
but do not commute with the operators $\hat L_m$,

\begin{equation}
[\hat L_n,\hat G_m] \varphi_R = m \alpha^{m+n} \varphi_R = - m\hat G_{m+n}\varphi_R.
\end{equation}

\noindent
Therefore, we obtain the algebra of the renormalization group in the case under consideration,

\begin{equation}\label{Commutations}
[\hat L_n,\hat L_m] = (n-m) \hat L_{n+m};\qquad [\hat G_n,\hat G_m] = 0;\qquad [\hat L_n,\hat G_m] = - m\hat G_{n+m},
\end{equation}

\noindent
where $n,m\ge 1$. The corresponding Jacobi identities can easily be verified, for example,

\begin{eqnarray}
&& [\hat L_m,[\hat L_n, \hat G_k]] + [\hat L_n,[\hat G_k,\hat L_m]] + [\hat G_k,[\hat L_m,\hat L_n]] = [\hat L_m,-k\hat G_{n+k}] + [\hat L_n,k\hat G_{m+k}] \nonumber\\
&& + [\hat G_k,(m-n)L_{m+n}] = \Big(k(n+k) - k(m+k) + (m-n)k\Big) \hat G_{m+n+k} = 0.
\end{eqnarray}

\noindent
Therefore, we really obtain the Lie algebra of the renormalization group for theories in which matter fields are also renormalized. Its commutation relations are given by Eq. (\ref{Commutations}).

\subsection{Representations in the presence of the matter renormalization}
\hspace*{\parindent}

To construct the explicit expressions for the renormalization group generators in the representation acting on the Green's functions, we use the perturbative expansion of the finite function $z_G(\alpha)$ present in Eq. (\ref{Green_Change}),

\begin{equation}
z_G(\alpha) \equiv z_G(\alpha,0) = 1 -\sum\limits_{n=1}^\infty (z_G)_n \alpha^n + O(a z_G, z_G^2),
\end{equation}

\noindent
where $(z_G)_n$ are small parameters, and $O(a z_G, z_G^2)$ includes all nonlinear terms in the Green's function renormalization. Then under the infinitesimal finite renormalization a Green's function changes as

\begin{equation}
\delta G_R = \sum\limits_{n=1}^\infty a_n \alpha^{n+1} \frac{\partial}{\partial\alpha} G_R - \sum\limits_{n=1}^\infty (z_G)_n \alpha^n G_R
\equiv \sum\limits_{n=1}^\infty \Big(a_n T_1(\hat L_n) + (z_G)_n T_1(\hat G_n)\Big) G_R.
\end{equation}

\noindent
Therefore, we obtain that in the considered representation the renormalization group generators are written in the form

\begin{equation}\label{Green_Representation_Full}
T_1(\hat L_n) = \alpha^{n+1} \frac{\partial}{\partial\alpha};\qquad T_1(\hat G_n) = -\alpha^n.
\end{equation}

\noindent
The commutation relations (\ref{Commutations}) are verified straightforwardly, e.g.,

\begin{eqnarray}
&& [T_1(\hat L_n),T_1(\hat G_m)] G_R = \Big(-T_1(\hat L_n) \alpha^m - T_1(\hat G_m) \alpha^{n+1}\frac{\partial}{\partial\alpha}\Big) G_R\nonumber\\
&& = -\alpha^{m+n+1} \frac{\partial}{\partial\alpha} G_R + \alpha^{n+1} \frac{\partial}{\partial\alpha}\Big(\alpha^m G_R \Big) = m \alpha^{m+n} G_R = - m T_1(\hat G_{m+n}) G_R.\qquad
\end{eqnarray}

Similarly, it is possible to construct explicit form of the considered generators in the representation acting on RGFs. For this purpose it is necessary to use the second equation in  (\ref{RGFs_Change}), which describes how the anomalous dimension changes under the finite renormalizations \cite{Vladimirov:1979ib}. Substituting into it the expressions (\ref{Alpha_Expansion}) for $\alpha'$ and (\ref{Z_Expansion}) for $z(\alpha)$ in the lowest (linear) order in $a_n$ and $z_n$ from this equation we obtain

\begin{eqnarray}\label{Delta_Gamma}
&& \gamma'(\alpha) -\gamma(\alpha) = \delta\gamma(\alpha) + O(a^2, az, z^2) = \sum\limits_{n=1}^\infty a_n\alpha^{n+1} \frac{d}{d\alpha} \gamma(\alpha) - \sum\limits_{n=1}^\infty z_n n \alpha^{n-1} \beta(\alpha)
\quad\nonumber\\
&& + O(a^2, az, z^2) \equiv \sum\limits_{n=1}^\infty \Big(a_n T_2(\hat L_n) + z_n T_2(\hat G_n)\Big) \gamma(\alpha) + O(a^2, az, z^2),\qquad
\end{eqnarray}

\noindent
where the infinitesimal change of the anomalous dimension $\delta\gamma$ by definition includes only terms linear in the parameters $a_n$ and $z_n$. The terms of higher orders in $a_n$ and $z_n$ in Eq. (\ref{Delta_Gamma}) are denoted by $O(a^2,az,z^2)$. From Eq. (\ref{Delta_Gamma}) and the analogous equation (\ref{Delta_Beta}) for the $\beta$-function we conclude that acting on RGFs the generators of the renormalization group give

\begin{eqnarray}
&& T_2(\hat G_n)\,\gamma(\alpha) = - n\alpha^{n-1}\beta(\alpha);\qquad\, T_2(\hat G_n)\,\beta(\alpha) = 0;\\
&& T_2(\hat L_n)\,\gamma(\alpha) = \alpha^{n+1} \frac{d\gamma(\alpha)}{d\alpha};\qquad\quad T_2(\hat L_n)\, \beta(\alpha) = \alpha^{n+1} \frac{d\beta(\alpha)}{d\alpha} - (n+1) \alpha^n\beta(\alpha).\nonumber
\end{eqnarray}

\noindent
It is convenient to present these equations in the matrix form,

\begin{eqnarray}\label{Explicit_L_And_G}
&& T_2(\hat L_n) \left(
\begin{array}{c}
\beta(\alpha)\vphantom{\displaystyle \frac{d}{d\alpha}}\\ \gamma(\alpha)\vphantom{\displaystyle \frac{d}{d\alpha}}
\end{array}
\right) = \left(
\begin{array}{cc}
{\displaystyle \alpha^{n+1}\frac{d}{d\alpha} - (n+1)\alpha^n} & 0\\
0 & {\displaystyle \alpha^{n+1}\frac{d}{d\alpha}}
\end{array}
\right) \left(
\begin{array}{c}
\beta(\alpha)\vphantom{\displaystyle \frac{d}{d\alpha}}\\ \gamma(\alpha)\vphantom{\displaystyle \frac{d}{d\alpha}}
\end{array}
\right);\qquad\nonumber\\
&&\vphantom{1}\nonumber\\
&& T_2(\hat G_n) \left(
\begin{array}{c}
\beta(\alpha)\vphantom{\displaystyle \frac{d}{d\alpha}}\\ \gamma(\alpha)\vphantom{\displaystyle \frac{d}{d\alpha}}
\end{array}
\right) = \left(
\begin{array}{cc}
0\vphantom{\displaystyle \frac{d}{d\alpha}} & 0\\
{\displaystyle -n\alpha^{n-1}} & 0\vphantom{\displaystyle \frac{d}{d\alpha}}
\end{array}
\right) \left(
\begin{array}{c}
\beta(\alpha)\vphantom{\displaystyle \frac{d}{d\alpha}}\\ \gamma(\alpha)\vphantom{\displaystyle \frac{d}{d\alpha}}
\end{array}
\right),
\end{eqnarray}

\noindent
which produces the explicit expressions for the renormalization group generators in the considered representation.

\subsection{Non-infinitesimal transformations}
\hspace*{\parindent}\label{Subsection_Finite}

Let us illustrate how one can obtain the (large) finite renormalizations involving the field renormalizations with the help of the exponential map. For this purpose we use the explicit expressions for the generators of the renormalization group given by Eq. (\ref{Explicit_L_And_G}) and present the transformation of RGFs in the form

\footnotesize
\begin{eqnarray}
&& \left(
\begin{array}{c}
{\displaystyle \beta'(\alpha)\vphantom{\Big(}}\\
{\displaystyle \gamma'(\alpha)\vphantom{\Big(}}
\end{array}
\right) = \exp\Big(\sum\limits_{n=1}^\infty a_n T_2(\hat L_n) + \sum\limits_{n=1}^\infty z_n T_2(\hat G_n)\Big)
\left(
\begin{array}{c}
{\displaystyle \beta(\alpha)\vphantom{\Big(}}\\
{\displaystyle \gamma(\alpha)\vphantom{\Big(}}
\end{array}
\right)
\nonumber\\
&& = \exp\Bigg\{
\left(
\begin{array}{cc}
{\displaystyle a_1 \Big(\alpha^2\frac{d}{d\alpha}-2\alpha\Big) + a_2\Big(\alpha^3\frac{d}{d\alpha}-3\alpha^2\Big)+\ldots} & {\displaystyle 0}\\
{\displaystyle -z_1 - 2z_2\alpha-3z_3\alpha^2 +\ldots} & {\displaystyle a_1 \alpha^2 \frac{d}{d\alpha} + a_2 \alpha^3 \frac{d}{d\alpha} + \ldots}
\end{array}
\right)
\Bigg\}
\left(
\begin{array}{c}
{\displaystyle \beta(\alpha)\vphantom{\Big(}}\\
{\displaystyle \gamma(\alpha)\vphantom{\Big(}}
\end{array}
\right)\nonumber\\
\end{eqnarray}
\normalsize

\noindent
The upper string gives the same result as obtained in Section \ref{Subsection_Finite_Alpha}, namely, the transformation of the $\beta$-function. Expanding the lower string in powers of $\alpha$ we derive the expression for the new anomalous dimension $\gamma'(\alpha)$ written in terms of the coefficients $a_n$ and $z_n$. However, it is more convenient to express the result in terms of the coefficients $k_n$ defined by Eq. (\ref{Alpha_Prime}) and similar coefficients $l_n$ which appear in the similar equation for $z(\alpha)$ (see Eq. (\ref{Z_Prime_Lowest}) below). This expression can again obtained with the help of the exponential map. For this purpose, we first write down Eq. (\ref{Finite_Z}) for the infinitesimal transformations with $\mu'=\mu$,

\begin{equation}
Z'(\alpha - \sum\limits_{n=1}^\infty a_n\alpha^{n+1},\ln\Lambda/\mu) = \Big(1-\sum\limits_{n=1}^\infty z_n \alpha^n \Big) Z(\alpha,\ln\Lambda/\mu) + O(a^2,az,z^2).
\end{equation}

\noindent
Then, the new renormalization constant for the matter fields in the linear approximation can be written as

\begin{eqnarray}
&& Z'(\alpha,\ln\Lambda/\mu) = \Big(1 + \sum\limits_{n=1}^\infty a_n\alpha^{n+1}\frac{\partial}{\partial\alpha}-\sum\limits_{n=1}^\infty z_n \alpha^n \Big) Z(\alpha,\ln\Lambda/\mu) + O(a^2,az,z^2)
\quad\nonumber\\
&& = \Big(1 -\sum\limits_{n=1}^\infty a_n\hat L_n - \sum\limits_{n=1}^\infty z_n\hat G_n + O(a^2,az,z^2)\Big) Z(\alpha,\ln\Lambda/\mu).\qquad
\end{eqnarray}

\noindent
Taking into account that the corresponding finite transformations form a group, it is possible to obtain the all-order expression for $Z'(\alpha,\ln\Lambda/\mu)$ with the help of the exponential map,

\begin{equation}
Z'(\alpha,\ln\Lambda/\mu) = \exp\Big(-\sum\limits_{n=1}^\infty a_n \hat L_n - \sum\limits_{n=1}^\infty z_n \hat G_n\Big) Z(\alpha,\ln\Lambda/\mu).
\end{equation}

\noindent
Next, we rewrite Eq. (\ref{Finite_Z}) (for the case $\mu'=\mu$) in the form

\begin{eqnarray}
&& z(\alpha) Z(\alpha,\ln\Lambda/\mu)= Z'(\alpha',\ln\Lambda/\mu) = \exp\Big(\sum\limits_{n=1}^\infty a_n \hat L_n \Big) Z'(\alpha,\ln\Lambda/\mu) \qquad\nonumber\\
&& = \exp\Big(\sum\limits_{n=1}^\infty a_n \hat L_n \Big) \exp\Big(-\sum\limits_{n=1}^\infty a_n \hat L_n - \sum\limits_{n=1}^\infty z_n \hat G_n\Big) Z(\alpha,\ln\Lambda/\mu).\qquad
\end{eqnarray}

\noindent
From this equation we conclude that the all-order expression for $z(\alpha)$ can be presented as

\begin{equation}\label{Z_Prime}
z(\alpha) = \exp\Big(\sum\limits_{n=1}^\infty a_n \hat L_n\Big) \exp\Big(-\sum\limits_{n=1}^\infty a_n \hat L_n - \sum\limits_{n=1}^\infty z_n \hat G_n\Big).
\end{equation}

\noindent
Using the Baker–Campbell–Hausdorff formula

\begin{equation}\label{BCH_Equation}
e^A e^B = \exp\Big(A+B+\frac{1}{2}[A,B] + \frac{1}{12}[A,[A,B]]+\frac{1}{12}[B,[B,A]] + \ldots\Big)
\end{equation}

\noindent
and the commutation relations (\ref{Commutations}) it is possible to verify that, in fact, the right hand side Eq. (\ref{Z_Prime}) does not contain any differential operators. For instance, in the lowest orders we obtain

\begin{eqnarray}\label{Z_Prime_Lowest}
&& z(\alpha) = 1 -\alpha z_1 + \alpha^2\Big(-z_2 +\frac{1}{2} z_1^2 +\frac{1}{2}a_1 z_1\Big) + \alpha^3\Big(-z_3 +z_1 z_2 -\frac{1}{6}z_1^3 + \frac{1}{2}a_2 z_1
\quad\nonumber\\
&& + a_1 z_2 - \frac{1}{3}a_1^2 z_1 -\frac{1}{2}a_1 z_1^2\Big) + O(\alpha^4) \equiv 1 + l_1 \alpha + l_2\alpha^2 + l_3\alpha^3 + O(\alpha^4).\qquad
\end{eqnarray}

\noindent
Using the relations between the coefficients $z_n$ and $l_n$ following from this equation we can derive how the anomalous dimension coefficients change under the considered finite renormalization,

\begin{eqnarray}\label{Gamma_Prime}
&& \gamma_1' = \gamma_1;\qquad \gamma_2' = \gamma_2 + \beta_1 l_1 -\gamma_1 k_1;\vphantom{\Big(}\nonumber\\
&& \gamma_3' = \gamma_3 + \beta_2 l_1 + \beta_1 (2l_2 - 2k_1 l_1 - l_1^2) - 2k_1 \gamma_2 + \gamma_1(2k_1^2 - k_2);\vphantom{\Big(}\nonumber\\
&& \gamma_4' = \gamma_4 + \beta_3 l_1 + \beta_2 (2l_2 - l_1^2-3k_1 l_1) + \beta_1 (3l_3-3l_2 l_1 + l_1^3 - 6k_1 l_2 + 3k_1 l_1^2 \quad\vphantom{\Big(}\nonumber\\
&& \quad\ \, + 5k_1^2 l_1 - 2k_2 l_1) - 3\gamma_3 k_1 + \gamma_2 (5k_1^2-2k_2) + \gamma_1(-k_3 - 5k_1^3 + 5k_2 k_1).\vphantom{\Big(}
\end{eqnarray}

\noindent
Exactly the same results are obtained directly from the second equation in (\ref{RGFs_Change}) after substituting into it the expressions for $\alpha'(\alpha)$ and $z(\alpha)$ given by Eqs. (\ref{Alpha_Prime}) and (\ref{Z_Prime_Lowest}).\footnote{In the particular case $l_1=0$, $\gamma_1=0$ Eqs. (\ref{Gamma_Prime}) exactly agree with the results of \cite{Kataev:1988sq} (after the replacements $\beta_n\to\beta_{n-1}$ and $\gamma_n\to\gamma_{n-2}$.)} Therefore, the argumentation based on the group structure of finite renormalizations perfectly agrees with the explicit equations (\ref{RGFs_Change}), at least, in the considered approximation.

\subsection{Rescaling subgroup}
\hspace*{\parindent}

Let us now consider finite renormalizations corresponding to the change of the renormalization point $\mu\to \mu'$, which now take the form

\begin{eqnarray}
&& \alpha(\mu) \to \alpha(\mu')\equiv \alpha'(\mu);\qquad\nonumber\\
&& Z\Big(\alpha(\mu),\ln\Lambda/\mu\Big) \to Z'\Big(\alpha(\mu'),\ln\Lambda/\mu'\Big) \equiv z\big(\alpha(\mu)\big) Z\Big(\alpha(\mu),\ln\Lambda/\mu\Big).\qquad
\end{eqnarray}

\noindent
Earlier we have demonstrated that the first transformation in the infinitesimal form are given by Eq. (\ref{Delta_Alpha_Rescaling}). Similarly, it is possible to rewrite the transformation for the field renormalization constant in the infinitesimal form,

\begin{eqnarray}
&& \varphi_R'-\varphi_R = - \delta z \varphi_R + O(t^2) = -\gamma(\alpha) t\, \varphi_R + O(t^2)\nonumber\\
&& = - t\sum\limits_{n=1}^\infty \gamma_n \alpha^n\, \varphi_R + O(t^2) = - t \sum\limits_{n=1}^\infty \gamma_n \hat G_n \varphi_R + O(t^2).\qquad
\end{eqnarray}

\noindent
Therefore, the overall transformation consists of both finite renormalizations of the coupling constant with $a_n=-t\beta_n$ and the finite renormalization of fields with the coefficients $z_n=-t\gamma_n$. This implies that they are generated by the operator

\begin{equation}
\hat L = - \sum\limits_{n=1}^\infty \Big(\beta_n \hat L_n + \gamma_n \hat G_n\Big).
\end{equation}

According to Eq. (\ref{Green_Representation_Full}), acting on the Green's functions this operator gives

\begin{eqnarray}
&& T_1(\hat L) G_R = - \sum\limits_{n=1}^\infty \Big(\beta_n T_1(\hat L_n) + (\gamma_G)_n T_1(\hat G_n)\Big) G_R\nonumber\\
&& = - \Big(\beta(\alpha) \frac{\partial}{\partial\alpha} - \gamma_G(\alpha)\Big) G_R = \frac{\partial}{\partial\ln\mu} G_R,
\end{eqnarray}

\noindent
where the last equality follows from the OCS equation (\ref{OCS}). Therefore, the considered transformations are again reduced to the shift of the renormalization point (certainly, as it should be). Evidently, they form the Abelian subgroup of the renormalization group.

Similarly, if we would like to consider the action of the operator $\hat L$ on RGFs, it is necessary to substitute the explicit expressions for the operators $\hat L_n$ and $\hat G_n$ given by Eq. (\ref{Explicit_L_And_G}). Then we obtain

\begin{eqnarray}
T_2(\hat L) \left(
\begin{array}{c}
\beta(\alpha)\vphantom{\displaystyle \frac{d}{d\alpha}}\\ \gamma(\alpha)\vphantom{\displaystyle \frac{d}{d\alpha}}
\end{array}
\right) = - \left(
\begin{array}{cc}
{\displaystyle \beta(\alpha)\frac{d}{d\alpha} - \frac{d\beta(\alpha)}{d\alpha}} & {\displaystyle 0\vphantom{\int\limits_p}} \\
{\displaystyle -\frac{d\gamma(\alpha)}{d\alpha}} & {\displaystyle \beta(\alpha)\frac{d}{d\alpha}}
\end{array}
\right) \left(
\begin{array}{c}
\beta(\alpha)\vphantom{\displaystyle \frac{d}{d\alpha}}\\ \gamma(\alpha)\vphantom{\displaystyle \frac{d}{d\alpha}}
\end{array}
\right) = 0.
\end{eqnarray}

\noindent
This implies that RGFs (i.e. $\beta(\alpha)$ and $\gamma(\alpha)$) remains invariant under the finite renormalizations (\ref{Mu_Change}) (which form a one-parametric Abelian subgroup of the renormalization group).

\section{Conclusion}
\hspace*{\parindent}

In this paper we analyzed the group structure of the renormalization group for theories with a single gauge coupling. In particular, we showed that the infinitesimal finite renormalizations of the gauge coupling constant form a subalgebra of the Witt algebra (the central extension of which is the well-known Virasoro algebra). The main difference of the algebra considered in this paper from the Witt (or Virasoro) algebra is that the index numerating the generators of the Witt algebra is an arbitrary integer, while the finite renormalizations of the coupling constant are numerated by a {\it positive} integer. Also, we constructed the commutation relations for the algebra of the infinitesimal finite renormalizations which also involve the matter field renormalizations. The finite renormalizations which belong to the corresponding Lie group can be obtained in standard way with the help of the exponential map. We also described the Abelian subgroup of the renormalization group which corresponds to the changing of the renormalization point. There are also some other subgroups corresponding to some special classes of renormalization schemes (see, e.g., \cite{Goriachuk:2018cac}), but we did not consider them in this paper.

The generators of the renormalization group were also presented in the explicit form in two different representations. In the first one they act on Green's functions, while in the second one they act on RGFs (namely, on the $\beta$-function and various anomalous dimensions). Certainly, the finite transformations in each representation can be obtained by the exponential map. In the Abelian case this can easily reproduce some equations desribing how the renormalization constants transform under the change of the renormalization scale.

Note that although in this paper we considered gauge theories, the results can be used for an arbitrary theory with a single coupling. It would be also interesting to generalize them to the case of theories with multiple couplings, which may possibly be done using the same technique.

\end{document}